\documentclass[fleqn,twoside]{article}
\usepackage{style}

\usepackage{graphicx}
\usepackage[figuresright]{rotating}
\usepackage{subfigure}

\graphicspath{{./images/}}

\input myunits.sty

\newcommand{\AmS}{{\protect\the\textfont2
  A\kern-.1667em\lower.5ex\hbox{M}\kern-.125emS}}

\title{First Results of the PixelGEM Central Tracking System for COMPASS}

\author{Alexander~Austregesilo\address[TUM]{Technische Universit\"at M\"unchen, Physik Department E18\\ James-Franck-Stra\ss e, 85748 Garching, Germany}\thanks{CERN, 1211\ Genève\ 23, Switzerland, phone:\ +41\,22\,7676466, mail:\ Alexander.Austregesilo@cern.ch\vspace{.3pc} We gratefully acknowledge the skills of Rui de Oliveira and his group at CERN TS-DEM-PMT, and the support of Christian Joram, Eric David, Miranda van Stenis, and Ian McGill of CERN PH-DT2. We are also grateful to Panasonic Electric Works Deutschland GmbH for providing us with samples of high-density connectors.
}
, Florian~Haas\addressmark, Bernhard~Ketzer\addressmark, Igor~Konorov\addressmark, Markus~Kr\"amer\addressmark, Alexander~Mann\addressmark, Thiemo~Nagel\addressmark, Stephan~Paul\addressmark and Sebastian~Uhl\addressmark}

\begin{document}
  
  \begin{abstract}
    For its physics program with a high-intensity hadron beam of up to
    $2\cdot 10^{7}\,\mathrm{particles}/\s$, the COMPASS experiment at CERN 
    requires tracking of charged particles scattered by very small angles
    with respect to the incident beam direction. While good resolution in time and space 
    is mandatory, the   
    challenge is imposed by the high beam intensity, requiring 
    radiation-hard detectors which add very
    little material to the beam path in order to minimize secondary
    interactions. 
    
    To this end, a set of triple-GEM detectors with a hybrid readout structure consisting of
    pixels in the beam region and 2-D strips in the periphery was designed and built.    
    Successful prototype tests proved the performance of this new detector type, showing both
    extraordinary high rate capability and detection efficiency. The amplitude information 
    allowed to achieve spatial resolutions about a factor of 10 smaller than the pitch and a 
    time resolution close to the theoretical limit imposed by the layout.

    The PixelGEM central tracking system consisting of five detectors, slightly improved with respect to the prototype,
    was completely installed in the COMPASS spectrometer in spring 2008.
    \vspace{1pc}
  \end{abstract}
  
  \maketitle
  
  \section{Introduction}
  
  COMPASS (COmmon Muon and Proton Apparatus for 
  Structure and Spectroscopy) is a two-stage magnetic
  spec\-tro\-me\-ter, built for the investigation of the gluonic and 
  quarkonic structure of nucleons and the spectroscopy of hadrons using high-intensity
  muon and hadron 
  beams 
  from CERN's Super Proton Synchrotron (SPS)
  \cite{Abbon:2007pq}. 
  
  Since spring 2008, hadron
  beams of up to $2\cdot 10^{7}\,\mathrm{particles}/\s$ are used to
  perform spectroscopy of mesons and baryons in the light quark sector. 
  These channels require the tracking of charged particles scattered by very
  small angles with respect to the incident beam, calling for  
  detectors with good resolution in space and time in order to disentangle pile-up and 
  multi-track events within the primary beam. Further demands of the high hadron
  flux density  are radiation hardness and minimal material budget in order to 
  avoid secondary interactions. 
  Since the beginning of the 2008 beam time, this task is performed by
  a set of five newly developed triple-GEM \cite{Altunbas:02a,Ketzer:04a} beam trackers with a hybrid
  readout structure, consisting of pixels in the central region and strips
  in two orthogonal projections in the periphery.
  Within the spectrometer, the so called PixelGEM detectors are grouped into three stations, 
  two of the latter consisting of two specimen which are rotated by $\pi/4$ against each other. 
  Owing to the redundancy given by the silicon vertex tracking detectors,
  the first station upstream of the first analyzing magnet
  consists of only one PixelGEM detector to minimize interactions there.

  \section{The PixelGEM detector}
  
\begin{figure}
    \centering
    \includegraphics[width=15pc]{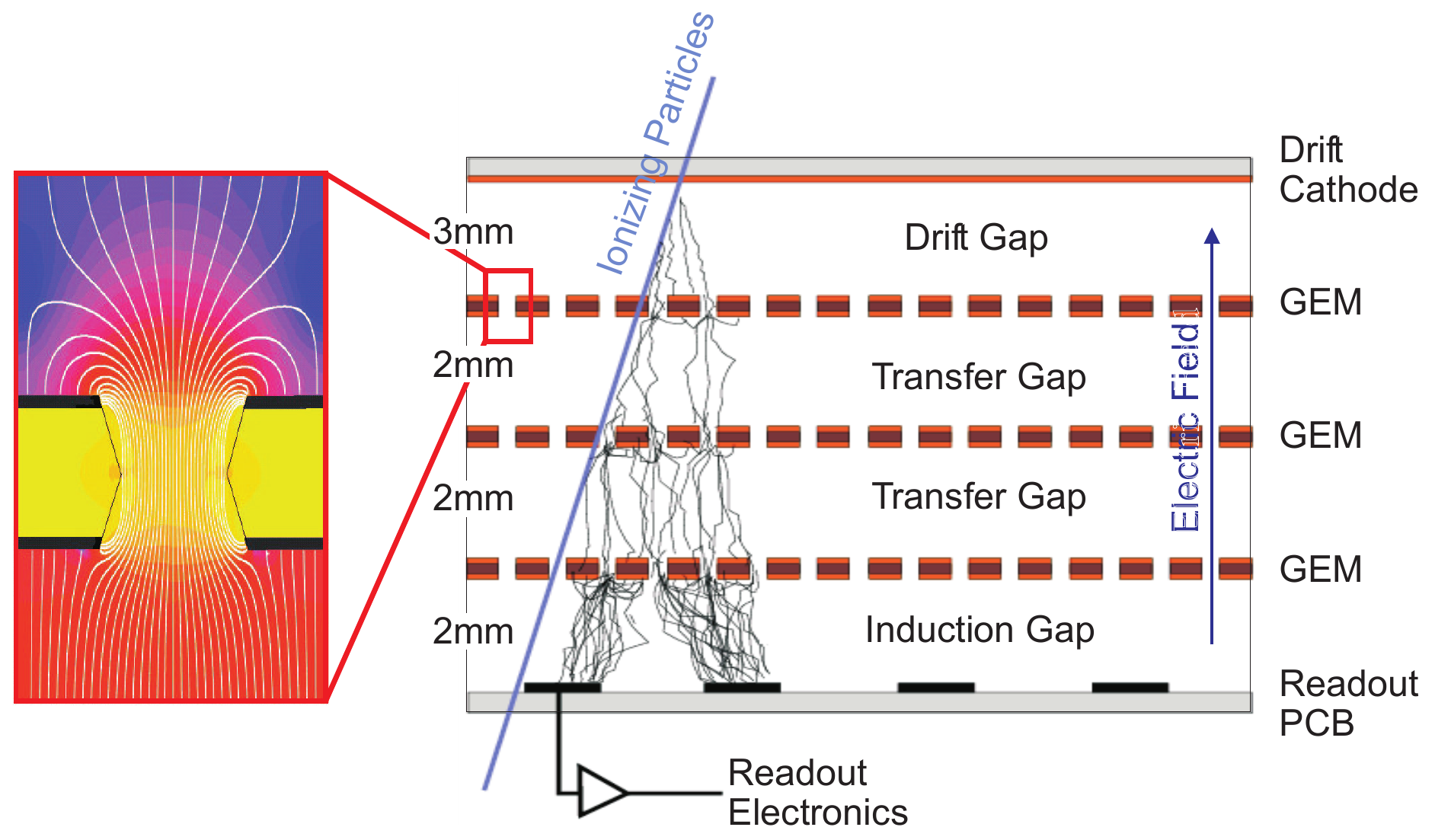}
    \caption{Schematic view of a GEM setup}
    \label{fig:GEM.geometry}
  \end{figure}

  The amplification stage of the PixelGEM detector uses the geometry
  already chosen for the COMPASS triple-GEM detectors with strip readout.
  This setup consists of a \(3\,\mathrm{mm}\) drift gap and three GEM foils, separated by \(2\,\mathrm{mm}\) transfer gaps
  from each other and by a \(2\,\mathrm{mm}\) induction gap from the readout circuit
  (c.f. Figure \ref{fig:GEM.geometry}).
  
  Avalanche charge amplification takes place in microscopic holes with a diameter of 
  \(70\,\mathrm{\mu m}\), which are placed in a hexagonal pattern with a pitch of \(140\,\mathrm{\mu m}\) 
  on a \(50\,\mathrm{\mu m}\) thick polyimide foil, coated with copper on both sides.
    
  The readout structure has been realized on a polyimide printed circuit foil
  of only \(100\,\mathrm{\mu m}\) total thickness with three conductive layers.
  The pixel size has been chosen to be \hbox{$1\!\times\!1\,\mm^2$}, which 
  constitutes a compromise between spatial resolution achievable and number of channels.
  Thus, the central area of \hbox{$32\!\times\!32\,\mm^2$} is covered by 1024 pixels.
  It is surrounded by a 2-D strip readout structure with a pitch of \hbox{$400\,\mum$}.
  In total, the active area of \hbox{$10\!\times\!10\,\cm^2$} is 
  covered using 2048 readout channels.  Analogue readout by the APV25-S1 ASIC in three sample mode has 
  been chosen in order to profit from amplitude measurements which help to improve 
  the spatial resolution by clustering neighboring hit strips or pixels.
  A detailed description of the clustering algorithm can be found in \cite{kraemer:08}.
  
  The total amount of material of the detector exposed to the beam corresponds
  to only 0.4\% of a radiation length $\mathrm{X}_0$ or 0.1\% of an interaction length
  $\lambda_{\mathrm{I}}$. 
  
\begin{figure}[h]
    \centering
    \includegraphics[width=10pc]{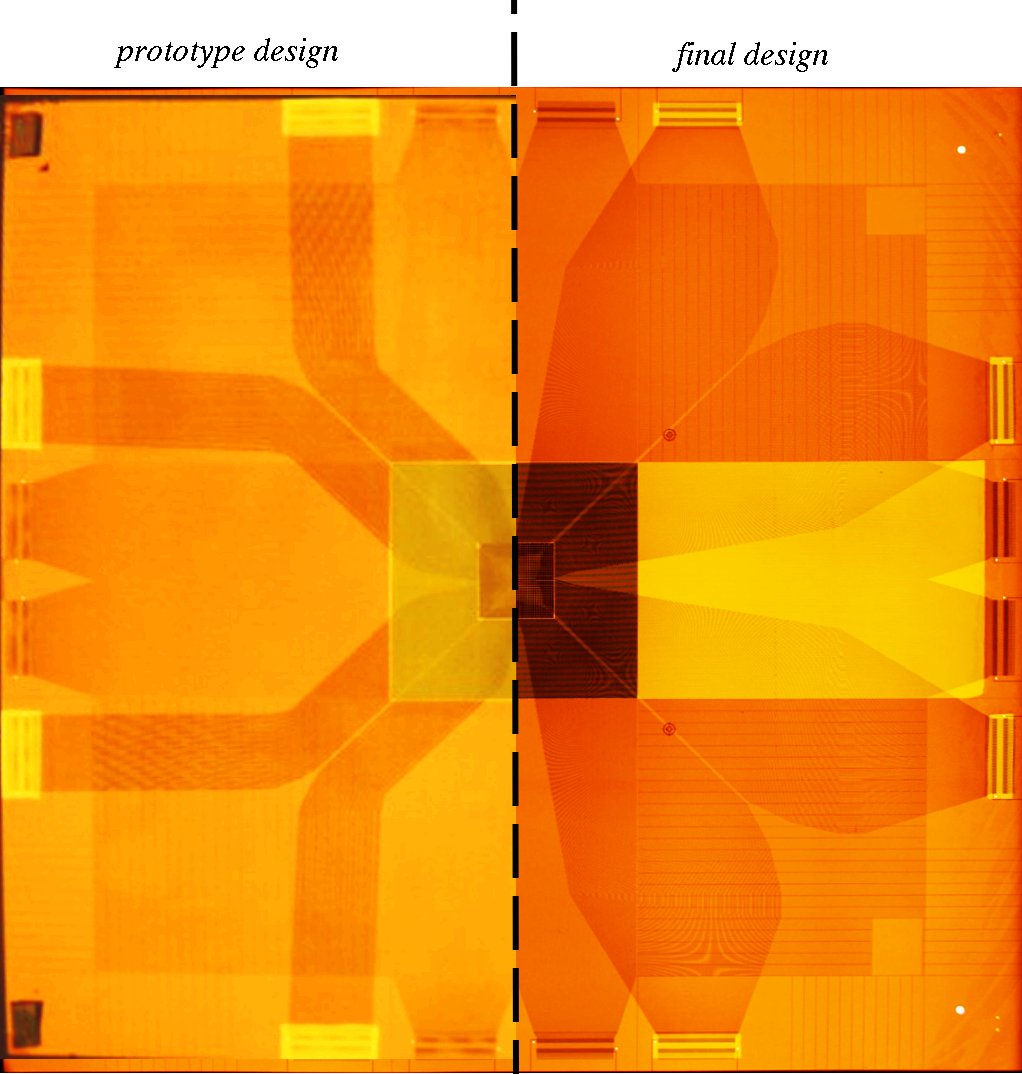}
    \caption{Photos of the readout printed circuit foils on a light table, 
      the prototype design (left hand side) and the final design (right hand side).}
    \label{fig:readout.lighttable}
  \end{figure}

  The signals from the active area are routed to front-end cards mounted at a
  distance of about $20\,\mathrm{cm}$ from the center by 
  long tracks on the printed circuit board (PCB). For the prototype,
  the small distance between these tracks of only about
  $200\,\mum$ caused crosstalk between neighboring channels in the order of \(15\%\) of the original signal.
  Although this crosstalk could be eliminated to a large extent by
  software, the routing of PCB tracks on the final readout foil was
  modified to maximize the distance between them. 
  Figure \ref{fig:readout.lighttable} shows the difference
  between the prototype and the final design
  of the readout foil. 
    
  The most significant design improvement concerns the 
  use of large area GEM and drift foils, which are coated with $1-2\,\mum$
  copper instead of $5\,\mum$. This reduces 
  the material budget of the detector by about 30\%, while gain and tracking
  performance were measured to be on the same level as for the prototype (see below).
  
  \section{Performance of the PixelGEM}
  
  In order to characterize the new detector both the prototype as well as
  a specimen with the final design were installed in front of the COMPASS
  silicon beam telescope upstream of the target for a few days in 2006 and 2007, 
  respectively. 
  The tracks used as reference for the analysis are measured by several silicon 
  and scintillating fiber planes and therefore have negligible spatial and temporal uncertainty
  with respect to the PixelGEM detector.
  
\begin{figure}[h]
    \centering
    \subfigure[\textit{muon beam}]{
      \includegraphics[angle=90, width=7.5pc]{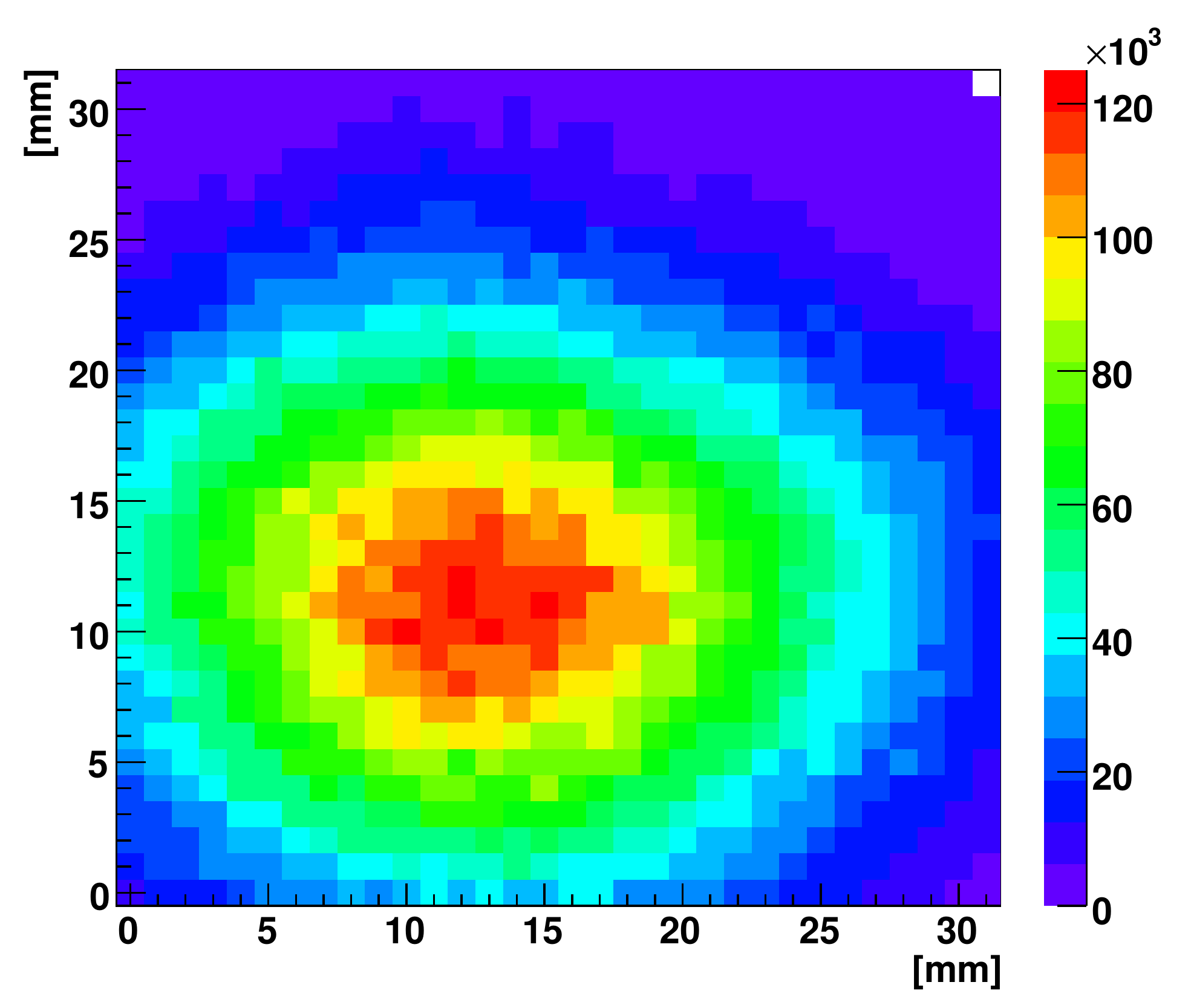}
    }
    \subfigure[\textit{hadron beam}]{
      \includegraphics[angle=90, width=7.5pc]{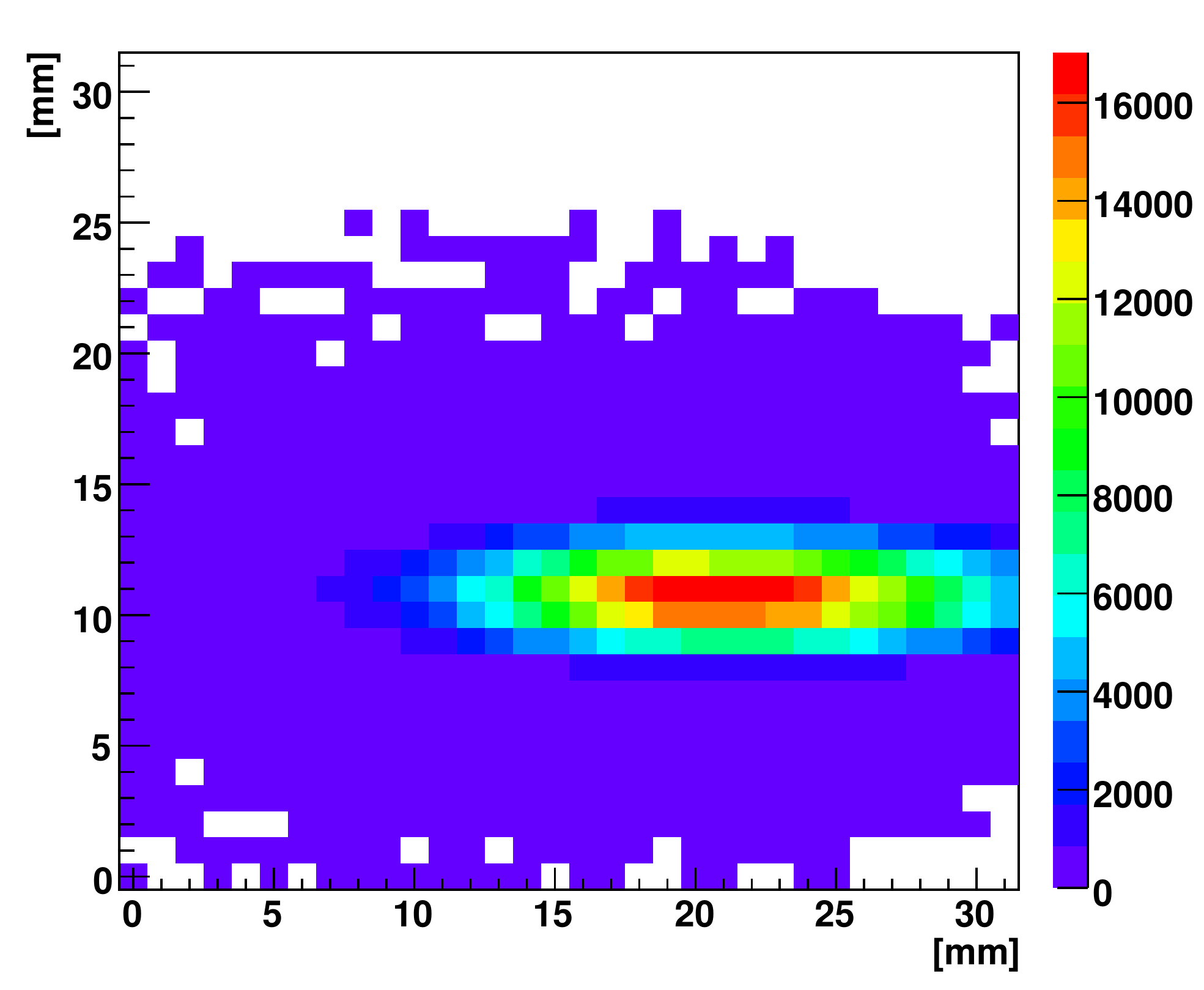}
    }
    \caption{Beam flux density \(\mathrm{ [ mm^{-2}s^{-1} ]}\) in the pixel region of the PixelGEM prototype (2006).}
    \label{fig:flux}
  \end{figure}

  To determine the performance of the PixelGEM, data recorded
  with muon and hadron beams were analyzed. The intensity
  and flux density of the $160\,\GeV/c$ muon beams reached up to 
  $4.8\cdot10^7/\s$ and $1.2\cdot10^5\,\mm^{-2}\s^{-1}$ \cite{aaust:07}, 
  respectively (c.f. Figure \ref{fig:flux} (a)). The intensity and flux density
  of the hadron beam was determined to be 
  $8.5\cdot10^6/\s$ and $1.6\cdot10^4\,\mm^{-2}\s^{-1}$ (c.f. Figure \ref{fig:flux} (b)).

  In the following, the most important performance characteristics of the pixel region of the 
  PixelGEM tracking detector
  will be discussed, comparing a \textit{low intensity} ($1.1\cdot10^6\,/\s$)
  and a \textit{high intensity} muon beam ($4.8\cdot10^7\,/\s$). 
  The performance of the detector in all analyzed hadron beams matches the one
  observed for low intensity muon beams and will not be discussed here. 
  A more detailed description of performance studies and assessment 
  under all measured beam conditions (including hadron beams) can 
  be found in \cite{aaust:07} and \cite{kraemer:08}.

  \subsection{Crosstalk suppression}
  
  Analyzing the beam test data, it turned out that the first design of the readout PCB leads to a significant 
  capacitive coupling between pixels, which are routed from the active
  area to the front-end electronics in adjacent tracks. Due to the complicated pattern,
  this so called crosstalk does not necessarily affect neighboring
  pixels.

\begin{figure}[h]
    \centering
    \subfigure[\textit{Without crosstalk suppression}]{
      \includegraphics[angle=90, width=6.5pc]{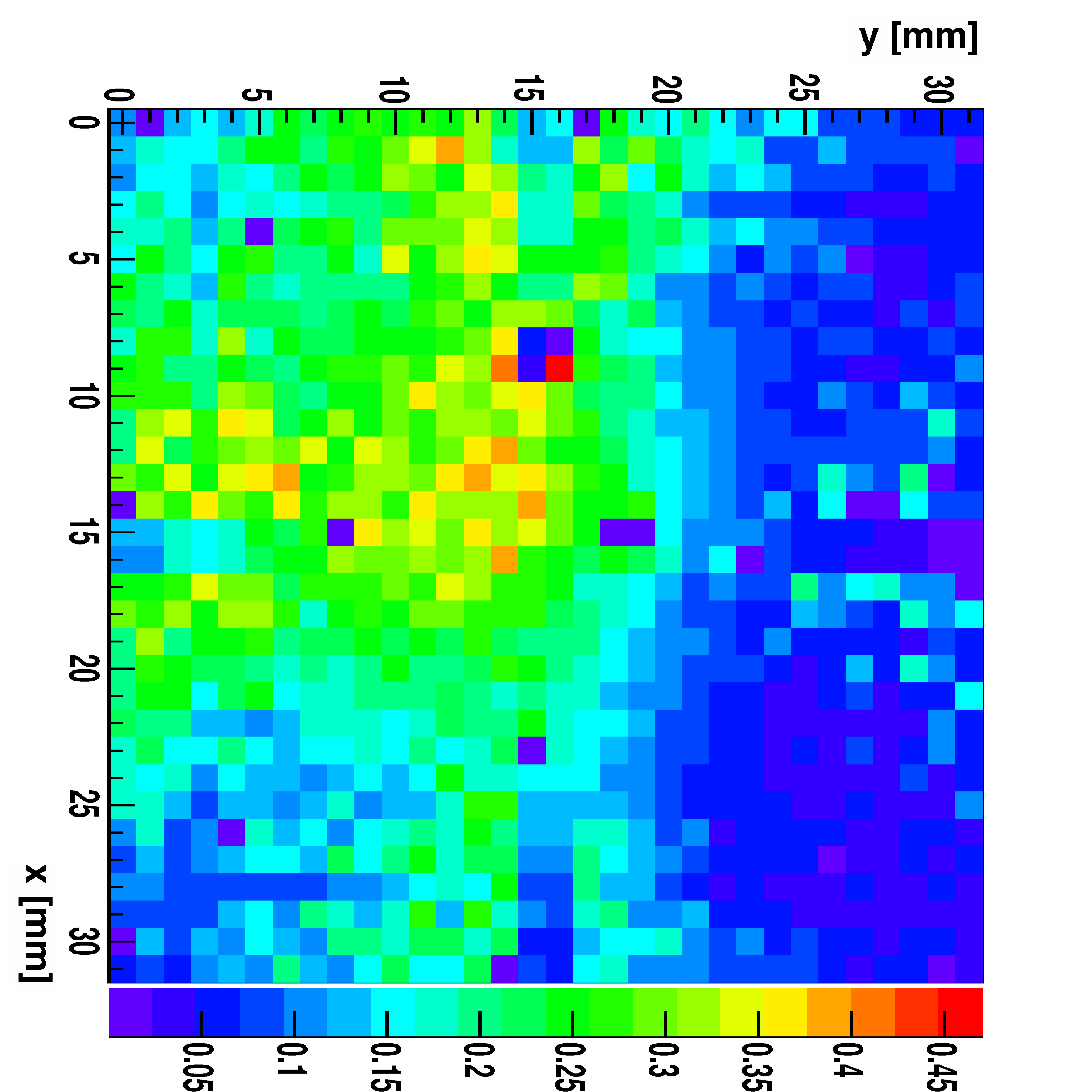}
    }
    \subfigure[\textit{After crosstalk suppression}]{
      \includegraphics[angle=90, width=6.5pc]{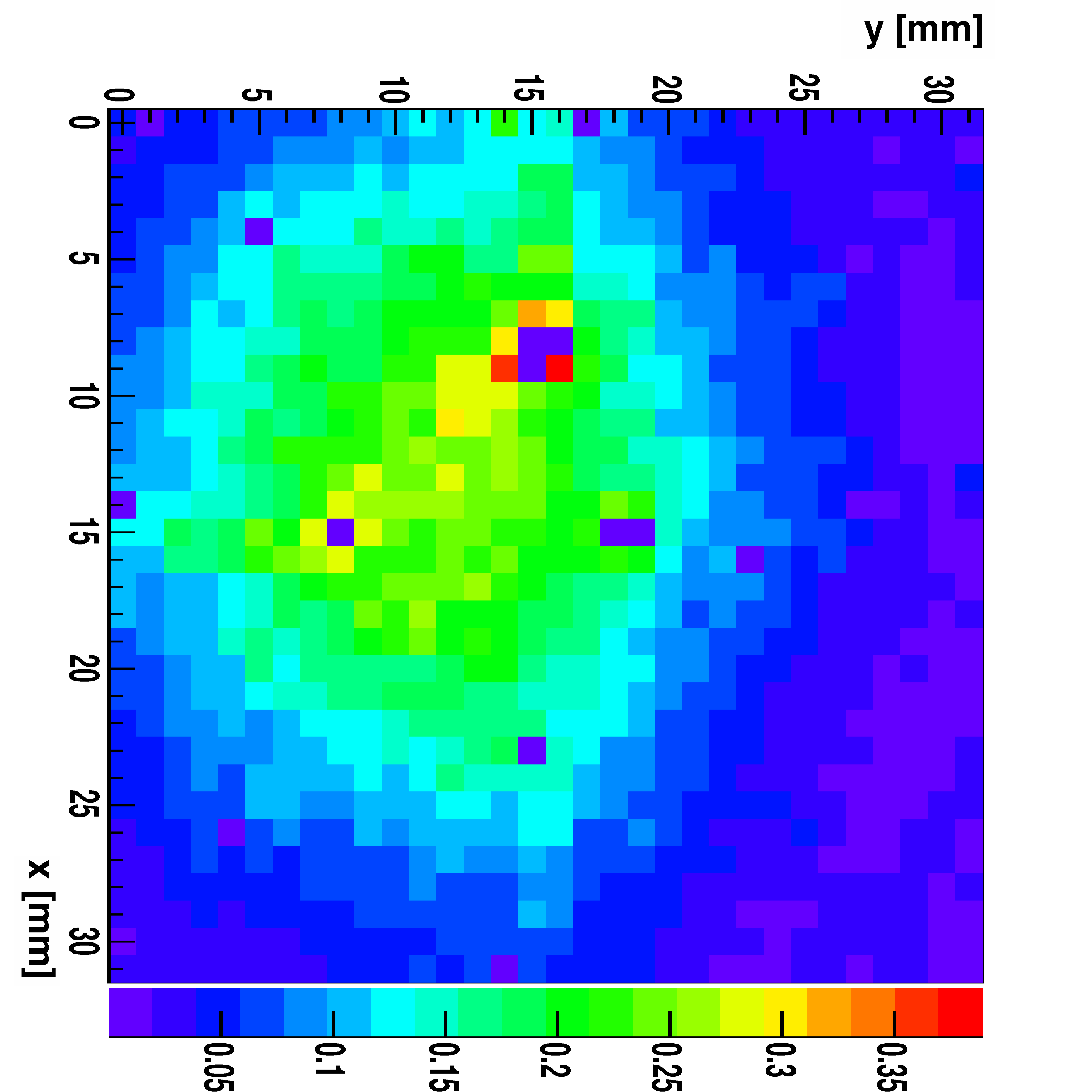}
    }
    \caption{Cluster position (pixel region) for low intensity muon beam. The color code is normalized to the
      number of events and given in percent.}
    \label{fig:crosstalk}
  \end{figure}

  In order to suppress this source of background during data processing, an algorithm was developed
  capable of eliminating crosstalk to are large extend (c.f. Figure \ref{fig:crosstalk}).

  The algorithm was tuned optimizing the performance of the detector as described in \cite{kraemer:08}. This tuning also indicates the improvement 
  of the final layout over the prototype readout design.

  \subsection{Efficiency}

\begin{figure}[h]
    \centering
    \subfigure[\textit{low intensity}]{
      \includegraphics[angle=90, width=8pc]{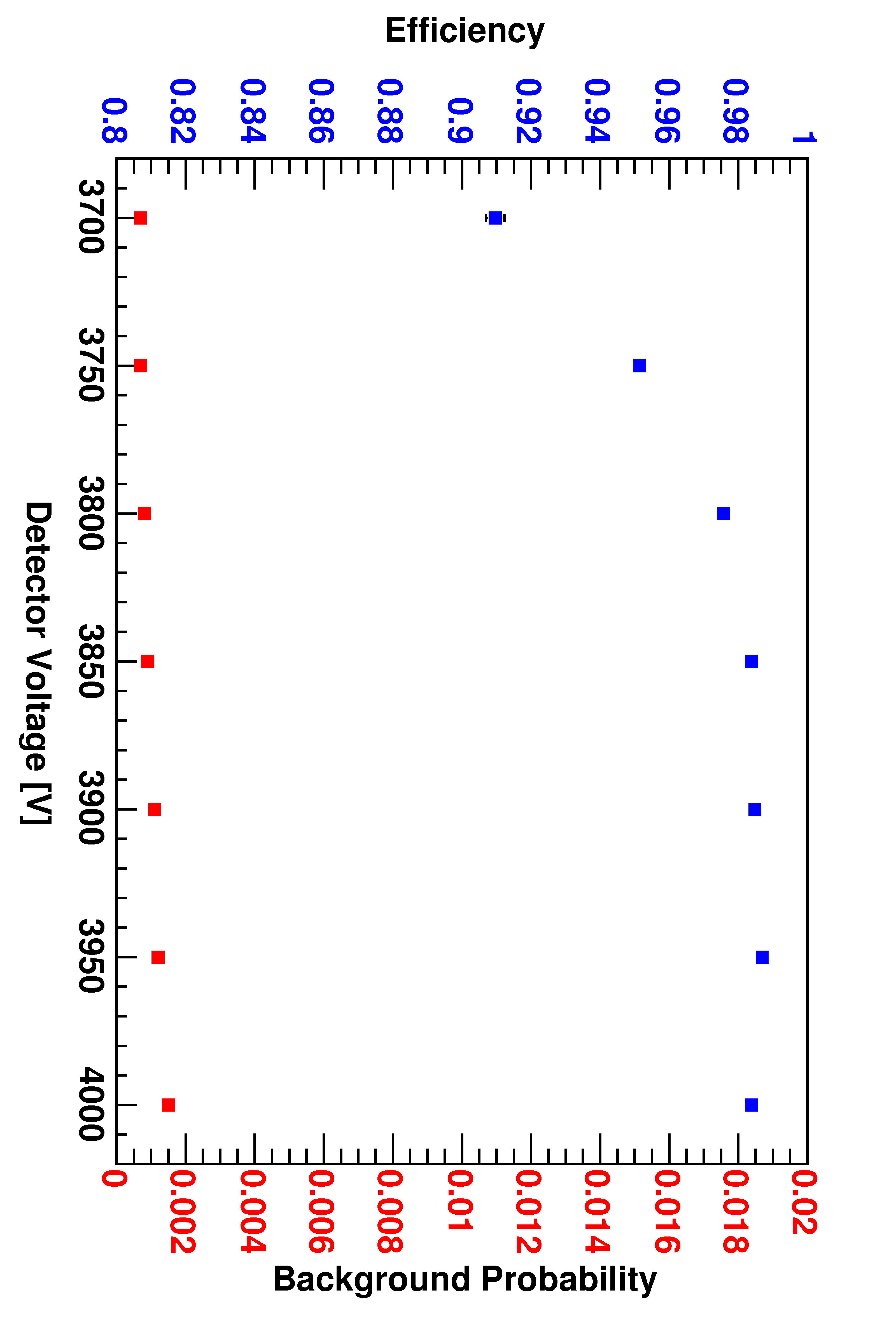}
    }
    \subfigure[\textit{high intensity}]{
      \includegraphics[angle=90, width=8pc]{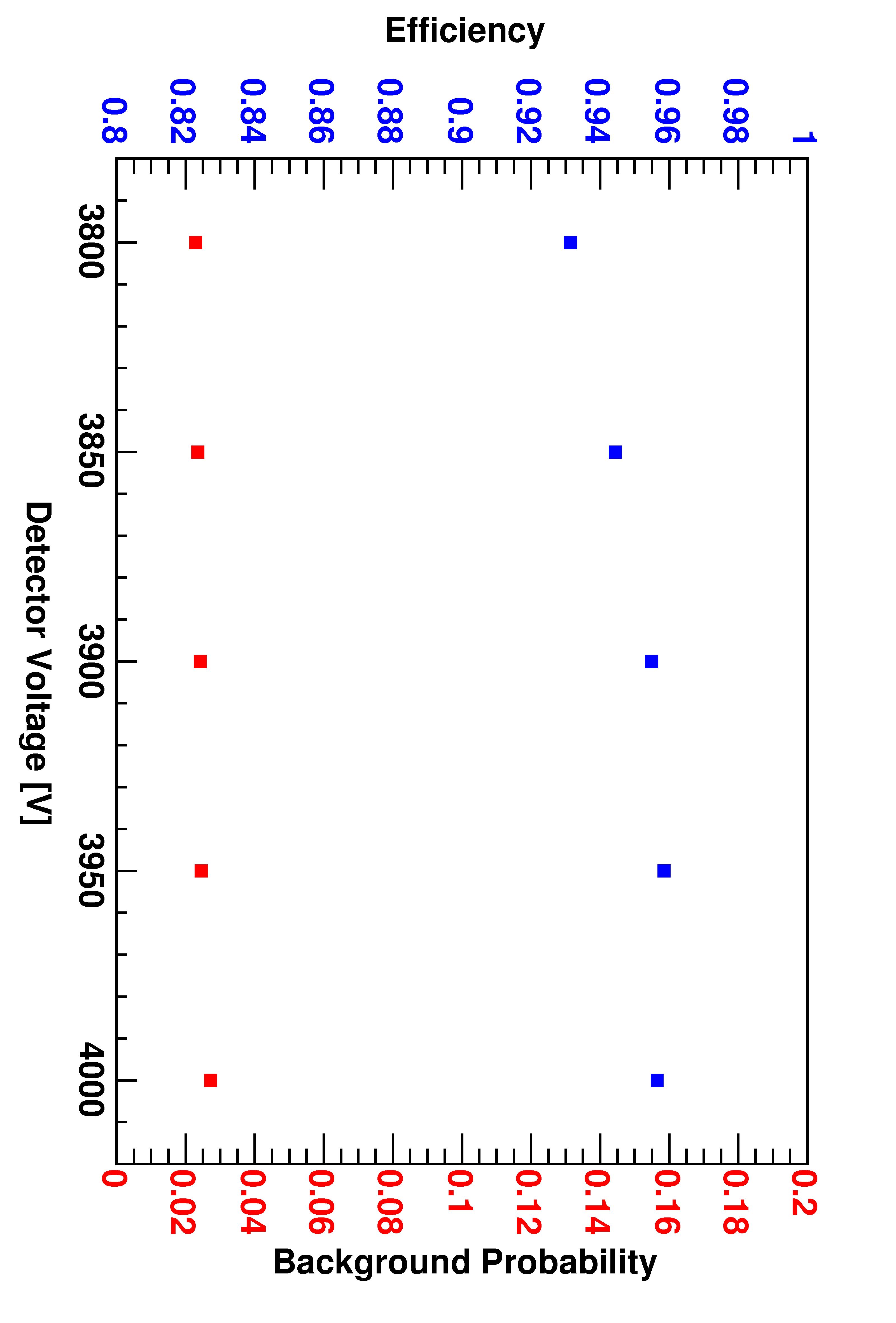}
    }
    \caption{Detection efficiency in the pixel region over high voltage settings for muon beam}
    \label{fig:eff.scan}
  \end{figure}

  To determine the nominal operation voltage for the detector a voltage scan 
  had been performed (c.f. Figures \ref{fig:eff.scan} (a) and  (b)). The nominal voltage 
  was chosen to be $3.9\,\kV$, where the plateau efficiency of $98.5\pm0.1\,\%$ for low and $96.4\pm0.1\,\%$
  for high intensity muon beams is reached. All further results mentioned in this publication 
  were obtained with this voltage.
  
  \subsection{Spatial resolution}

\begin{figure}[h]
    \centering
    \subfigure[\textit{low intensity}]{
      \includegraphics[angle=90, width=8pc]{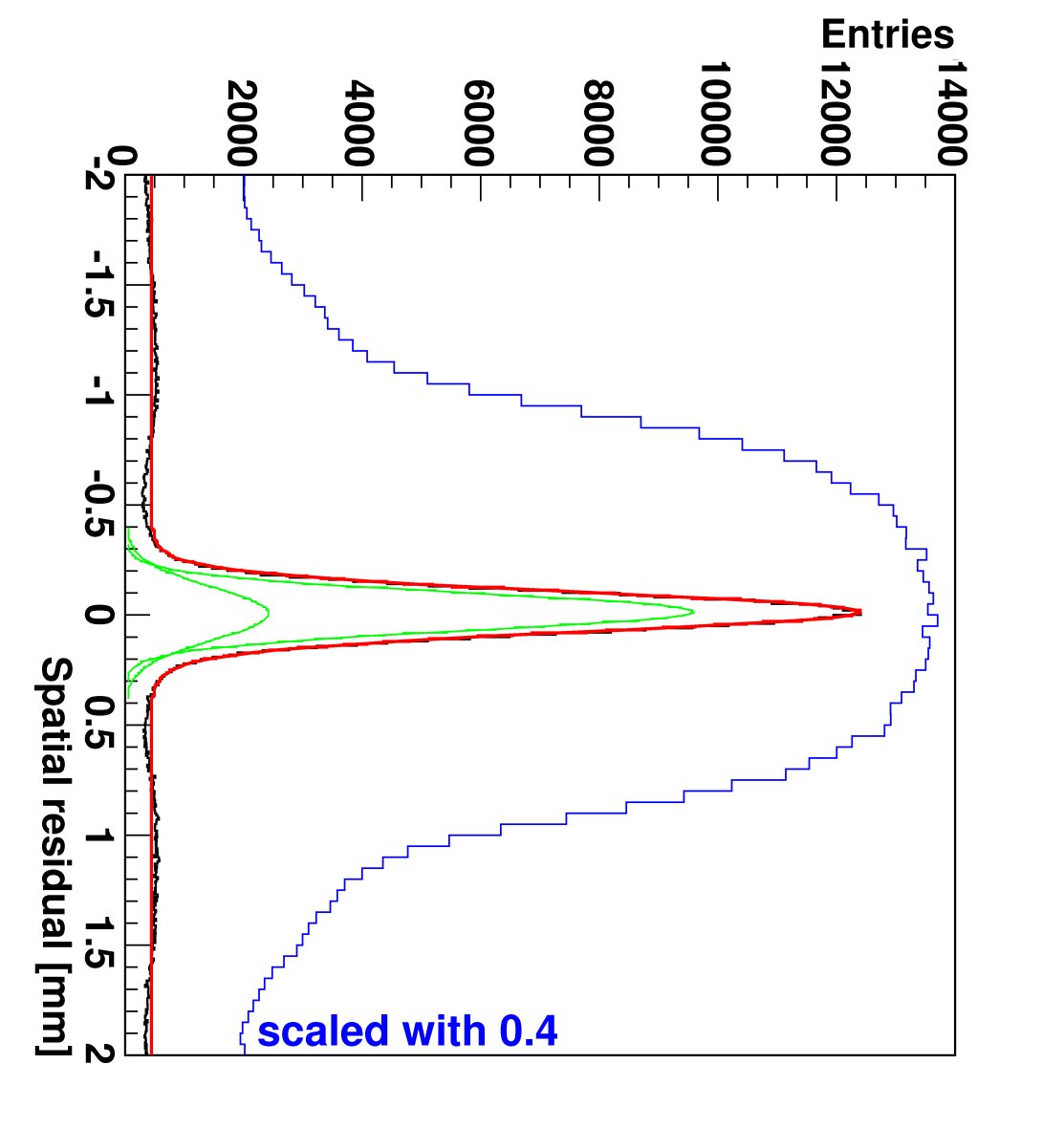}
    }
    \subfigure[\textit{high intensity}]{
      \includegraphics[angle=90, width=8pc]{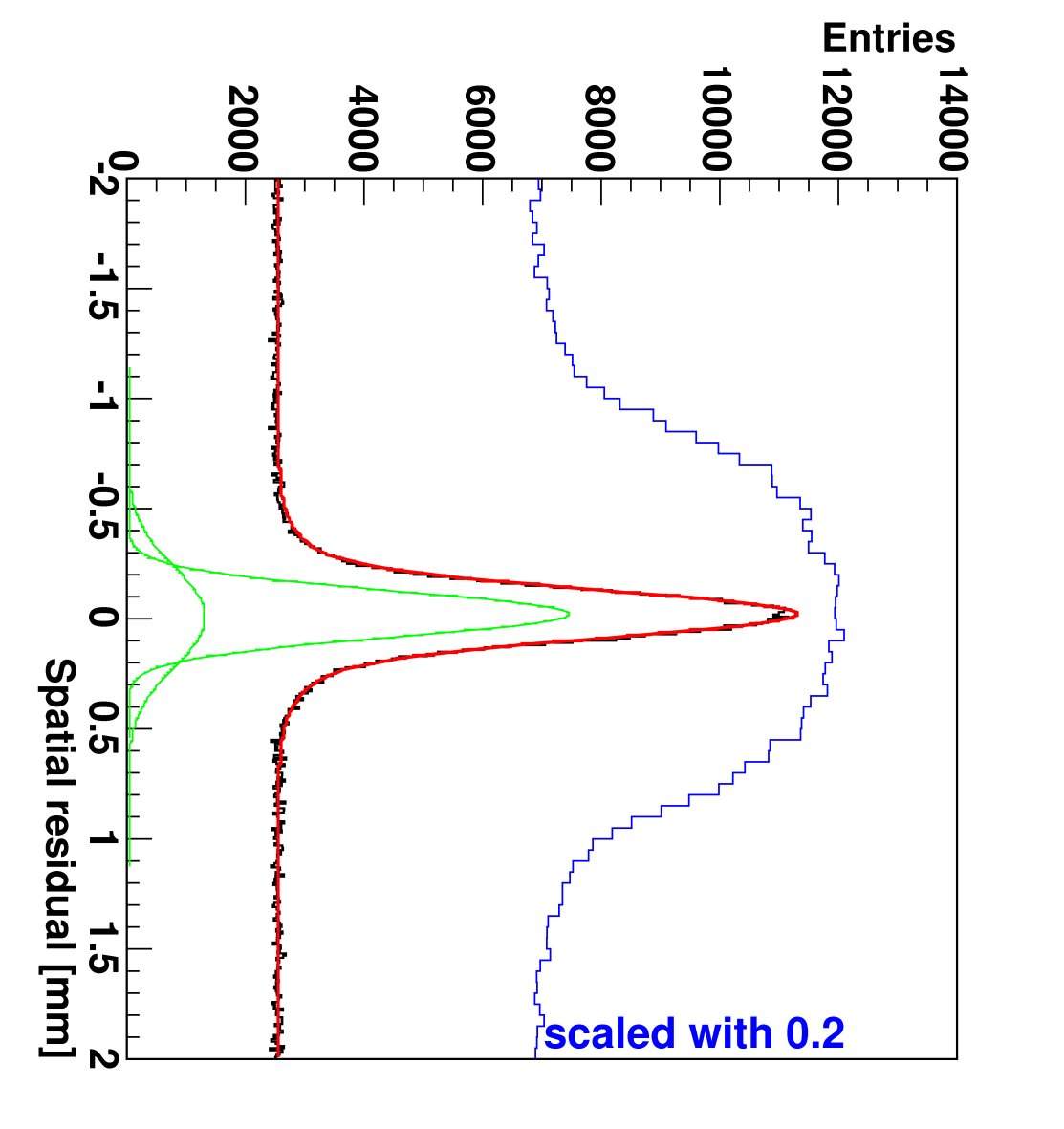}
    }
    \caption{Residual distribution in the pixel region (blue: raw hits, black: clusters, red: fit, green: Gaussian components of the fit)}
    \label{fig:res}
  \end{figure}

  By fitting a double Gaussian function with flat
  background to the residual distributions (c.f. Figures \ref{fig:res} (a) and (b)),
  the spatial resolution can be determined as the weighted mean of 
  the two components' standard deviations.
  
  The spatial resolution achieved for low intensity muon beams is \(90\,\mathrm{\mu m}\),
  while for high intensity muon beams the spatial resolution is increased to \(135\,\mathrm{\mu m}\).
  The reason for the decreased performance under high intensity beam conditions is mostly due to
  pile-up of off-time tracks.
  
  \subsection{Temporal resolution}
  
\begin{figure}[h]
    \centering
    \subfigure[\textit{low intensity}]{
      \includegraphics[angle=90, width=7pc]{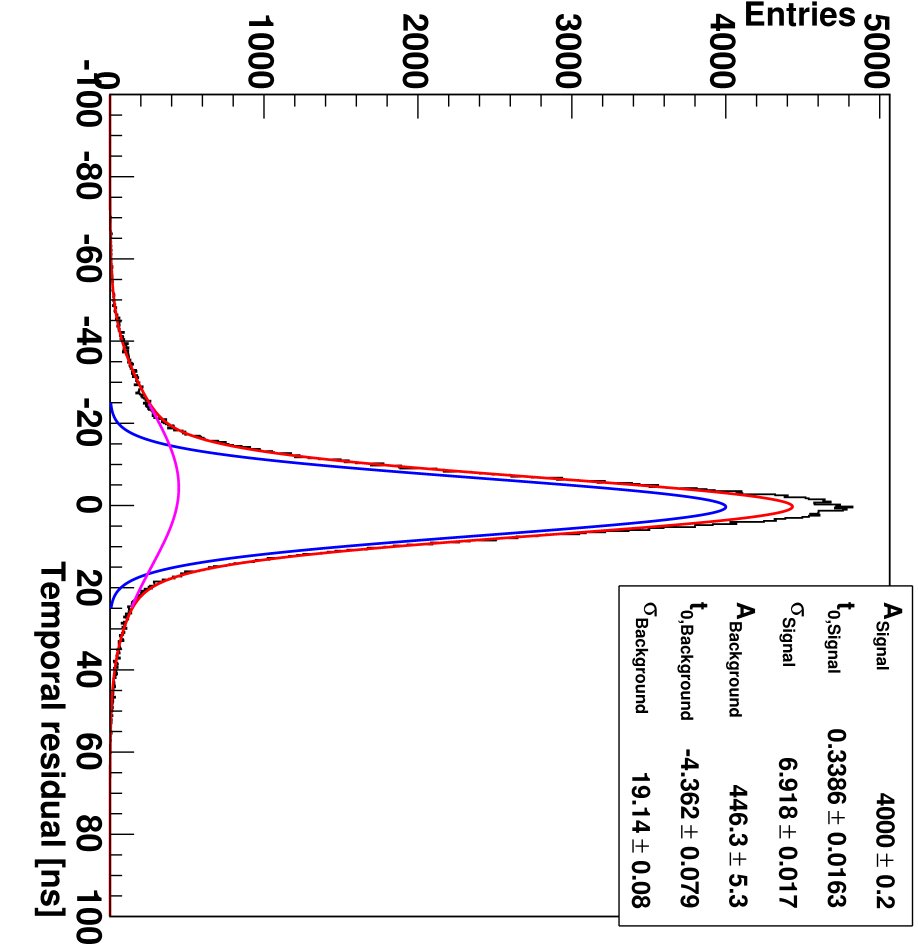}
    }
    \subfigure[\textit{high intensity}]{
    \includegraphics[angle=90, width=7pc]{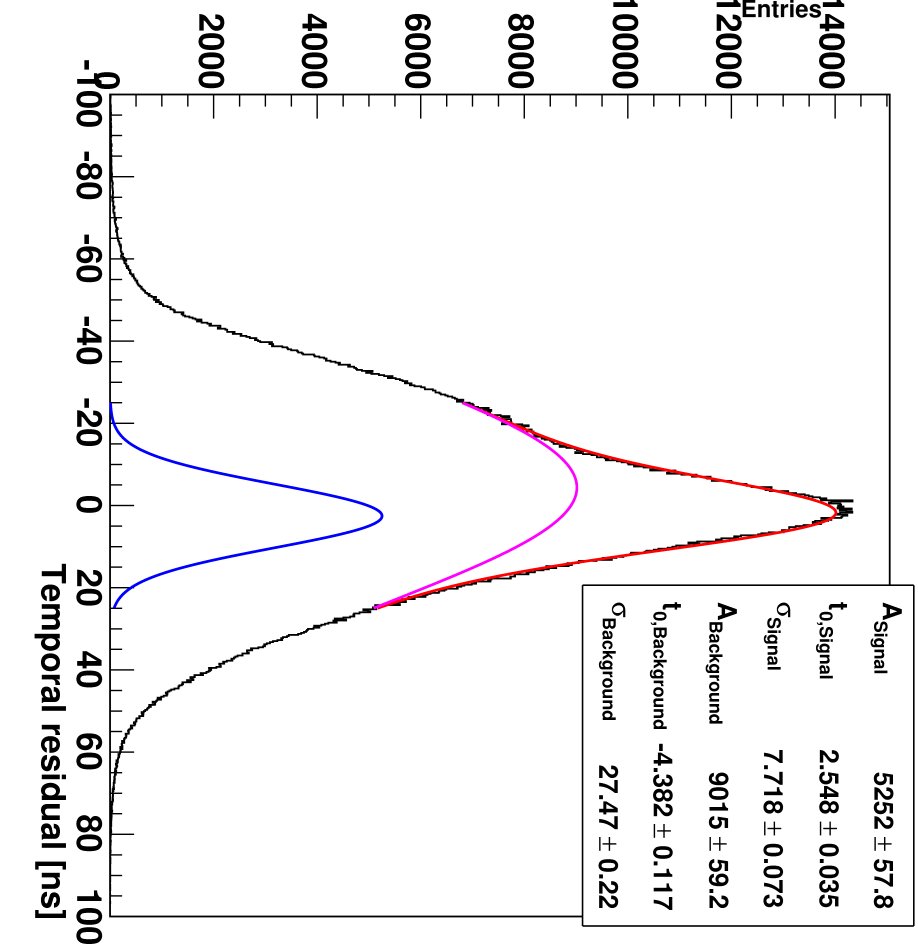}
    }
    \caption{Temporal residual distribution of pixel clusters (red: fit, blue: signal, magenta: background)}
    \label{fig:tres}
  \end{figure}

  The analogue readout of the PixelGEM detector does not allow for a direct time measurement.
  Thus the track time has to be determined applying pulse shape analysis to the three amplitude samples,
  which are read out for each event \cite{kraemer:08}.
  Figure \ref{fig:tres} (a) and (b) show the temporal residual distribution for
  both low and high intensity muon beams.
  The temporal resolution is determined to
  be below $8\,\ns$ for both cases. This can be confirmed off-line, 
  by imposing spatial track selection to suppress the background.
  
  \section{Conclusion}
  
  Being adapted to the special requirements of the hadron program of COMPASS,
  the novel PixelGEM detector met the design performance during the beam test in 2006
  and 2007. In advance of the 2008 hadron run, the full set of five detectors was 
  installed in the COMPASS spectrometer. Results from the ongoing analysis 
  of the complete PixelGEM central tracking system performance will be available soon.

\end{document}